\documentclass[acmtog]{acmart}

\usepackage{array}
\usepackage{gensymb}
\usepackage{multirow}

\usepackage{hyperref}

\usepackage{lipsum}

\DeclareMathOperator*{\argmin}{arg\,min}

\usepackage{atbegshi}
\newcommand{\placetextbox}[3]{
  \setbox0=\hbox{#3}
  \AtBeginShipoutNext{\AtBeginShipoutUpperLeft{%
    \put(\dimexpr#1\paperwidth\relax,-\dimexpr#2\paperheight\relax)
    {\vtop{{\null}\makebox[0pt][c]{#3}}}%
  }}%
}

\AtBeginDocument{%
  \providecommand\BibTeX{{%
    \normalfont B\kern-0.5em{\scshape i\kern-0.25em b}\kern-0.8em\TeX}}}

\acmSubmissionID{582}

\newcommand{\revision}[1]{{#1}}

\begin{document}

\copyrightyear{2025}
\acmYear{2025}
\setcopyright{acmlicensed}
\acmConference[SIGGRAPH Conference Papers '25]{Special Interest Group on Computer Graphics and Interactive Techniques Conference Conference Papers }{August 10--14, 2025}{Vancouver, BC, Canada}
\acmBooktitle{Special Interest Group on Computer Graphics and Interactive Techniques Conference Conference Papers (SIGGRAPH Conference Papers '25), August 10--14, 2025, Vancouver, BC, Canada}
\acmDOI{10.1145/3721238.3730666}
\acmISBN{979-8-4007-1540-2/2025/08}

\title{Physically Controllable Relighting of Photographs}

\author{Chris Careaga}

\author{Ya\u{g}{\i}z Aksoy}
\affiliation{
  \institution{Simon Fraser University}
  \city{Burnaby}
  \state{BC}
  \country{Canada}
}

\begin{abstract}
We present a self-supervised approach to in-the-wild image relighting that enables fully controllable, physically based illumination editing.
We achieve this by combining the physical accuracy of traditional rendering with the photorealistic appearance made possible by neural rendering.
Our pipeline works by inferring a colored mesh representation of a given scene using monocular estimates of geometry and intrinsic components.
This representation allows users to define their desired illumination configuration in 3D. The scene under the new lighting can then be rendered using a path-tracing engine.
We send this approximate rendering of the scene through a feed-forward neural renderer to predict the final photorealistic relighting result.
We develop a differentiable rendering process to reconstruct in-the-wild scene illumination, enabling self-supervised training of our neural renderer on raw image collections.
Our method represents a significant step in bringing the explicit physical control over lights available in typical 3D computer graphics tools, such as Blender, to in-the-wild relighting.



\end{abstract}



\begin{teaserfigure}
\centering
  \includegraphics[width=\linewidth]{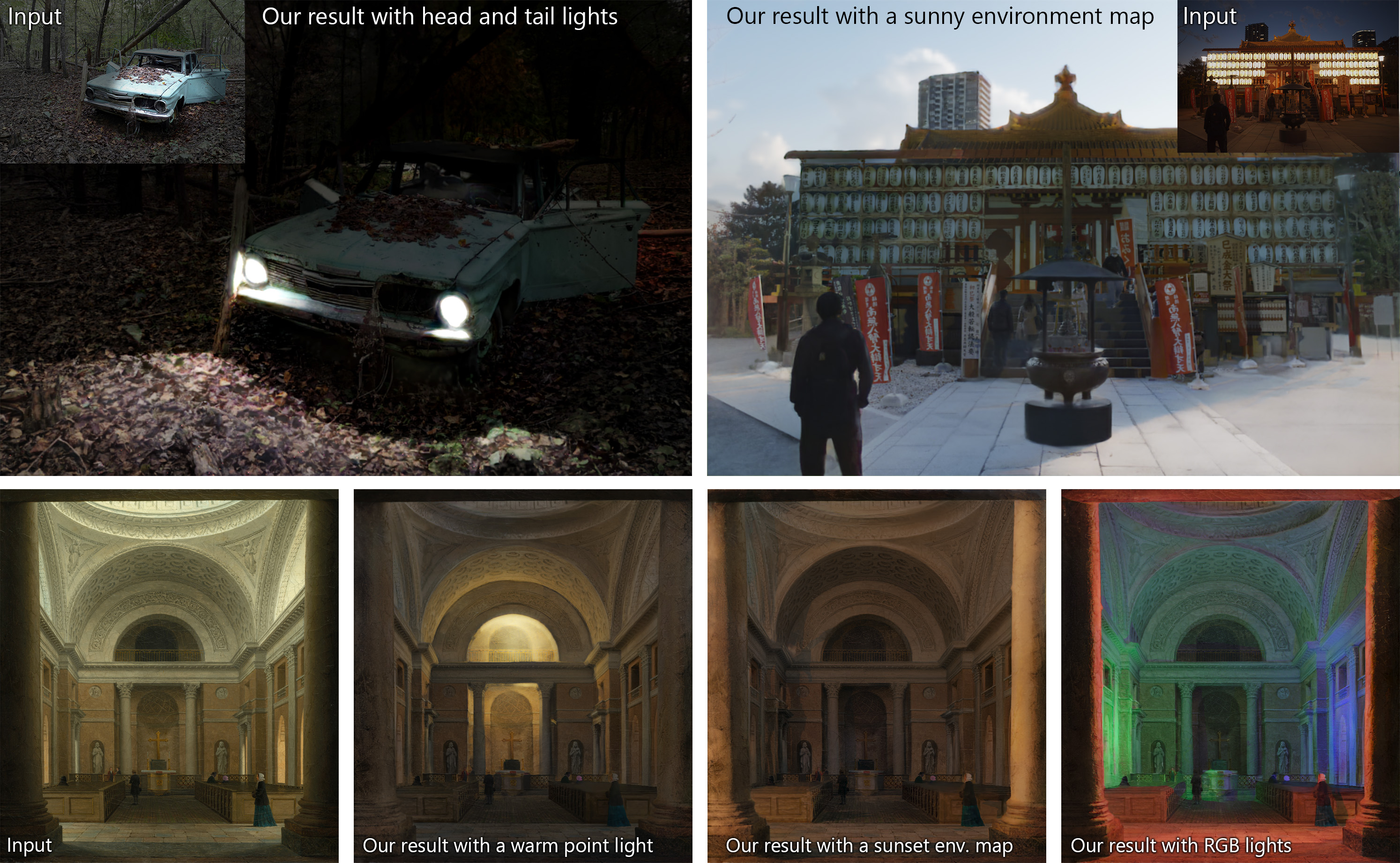}
  \vspace{-0.4cm}
  \caption{We present a photograph relighting method that enables explicit control over light sources akin to CG pipelines. 
  As the examples show, users can insert different types of light sources, such as spot lights, point lights, or environmental illumination into the scene. We achieve this in a pipeline involving mid-level computer vision, physically-based rendering, and neural rendering. 
  We introduce a self-supervised training methodology using differentiable rendering to train our neural renderer with real-world photograph collections for in-the-wild generalization. \mbox{} \hfill \footnotesize{Images from Unsplash by Patti Black (car) and Heinrich Hansen - courtesy of The Cleveland Museum of Art (church).}}
  \label{fig:teaser}
\end{teaserfigure}

\maketitle

\placetextbox{0.14}{0.03}{\includegraphics[width=4cm]{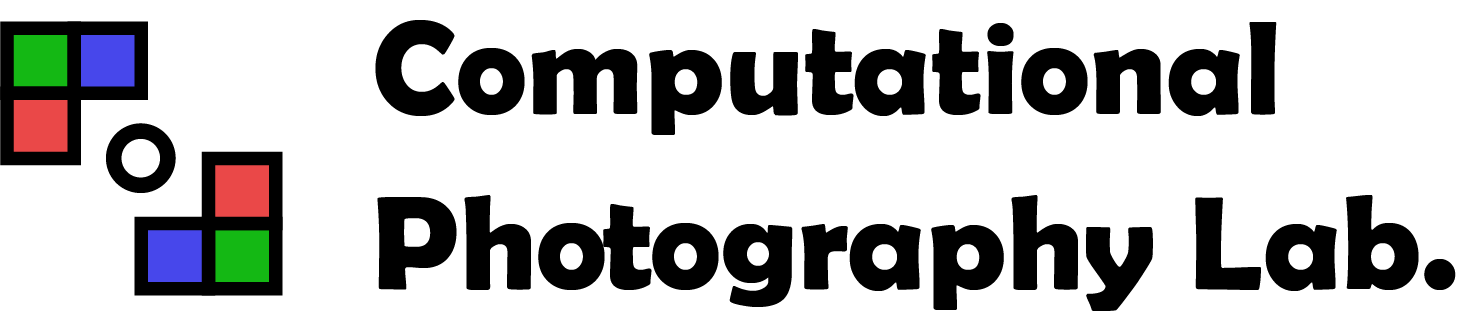}}
\placetextbox{0.85}{0.03}{Find the project web page here:}
\placetextbox{0.85}{0.045}{\textcolor{purple}{\url{https://yaksoy.github.io/PhysicalRelighting/}}}

\section{Introduction}
\label{sec:intro}
Illumination is an essential component of image formation. 
Computer graphics pipelines, more specifically physically-based rendering (PBR) engines such as Blender \cite{blender} and Unreal Engine \cite{unrealengine}, offer full control over light sources in the 3D scene and use ray tracing to model the interactions between illumination and geometry. 
With the ability to simulate different types of light sources, such as environmental, diffuse, directional, projected, or point, they allow unconstrained control over the scene illumination in a CG environment.

Although image relighting has been widely studied in the literature, computational photography methods have consistently failed to offer a similar unconstrained control over illumination for in-the-wild photographs. 
This stems from the complexity of modeling the interactions between illumination and the 3D scene in real-world image formation and the lack of real-world ground truth. 
As a result, prior interactive image relighting methods use simplified illumination conditions such as an HDRI environment \cite{sengupta19neural} or offer indirect control over light through scribbles \cite{choi2024scribble}, using a second image for light conditioning \cite{xing2024luminet}, or text-based descriptions \cite{zeng2024rgbx}.

\begin{figure*}[ht]
  \includegraphics[width=\linewidth]{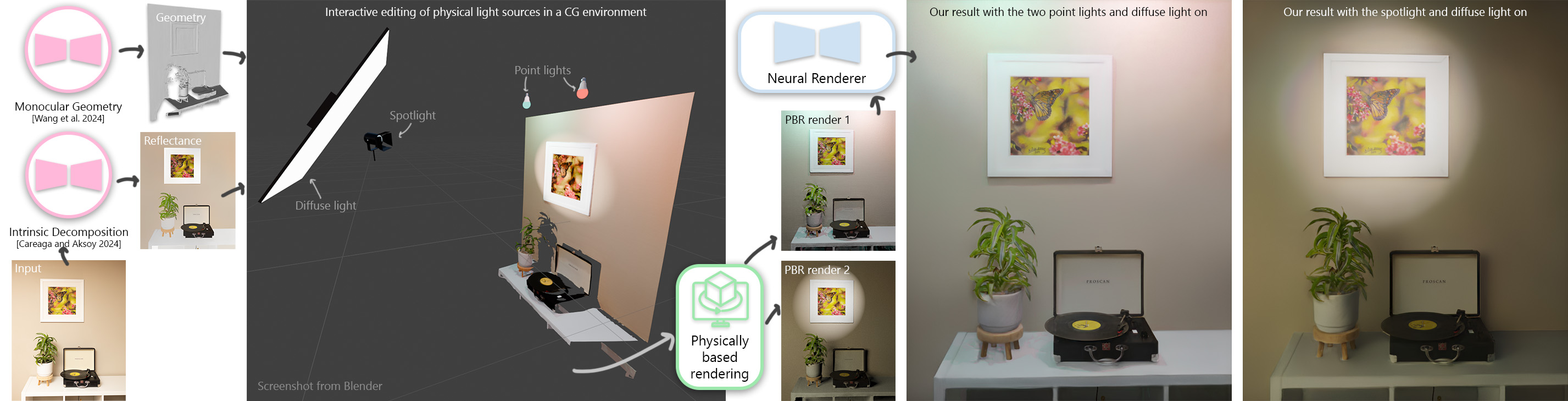}
  \caption{During inference, our pipeline first extracts PBR quantities using two off-the-shelf mid-level vision estimators (CID and MoGe). We use these representations to build a colored mesh of the scene. The user can then create their desired lighting environment in 3D editing software like Blender. We use Monte-Carlo rendering to create a rough approximation of the scene and run it through our self-supervised neural renderer to generate the final result. }
  \label{fig:pipeline_fwd}
\end{figure*}

In this work, we bring the capability of physical control over light sources to the relighting of in-the-wild photographs. 
We achieve this by combining a physically based rendering engine with a feed-forward neural renderer. 
Given an input image, our pipeline starts with monocular geometry estimation \cite{wang2024moge} and intrinsic decomposition \cite{careaga2024colorful} that together allow us to partially recreate the scene as a textured mesh in a 3D rendering environment. 
This step enables full control over the illumination in the scene where the user can freely define light sources or environmental illumination using interactive tools such as Blender or frameworks such as Mitsuba \cite{jakob2022mitsuba3}. 
Our system first renders an initial image through ray tracing using the customized scene illumination.
Given this initial CG rendering, our neural renderer generates the final photorealistic result. 
Figure~\ref{fig:pipeline_fwd} shows an overview.
\revision{While it is possible to render an approximate rendition of the scene under a given illumination, creating a photorealistic relighting requires a neural renderer that can model real-world appearance under custom illuminations.}
While generating training pairs under known different illumination conditions is possible using 3D assets, synthetic datasets fail to reflect the complex appearance and variety of real-world photographs.  
Real-world capture of paired data requires expensive photography systems in studio conditions, failing to provide the wide variety of scenes and illumination conditions required for in-the-wild generalization.

Our main contribution in this paper is the self-supervised training strategy we develop for our neural renderer. 
We introduce an optimization formulation that can replicate the existing illumination in a given image in a PBR environment using differentiable rendering. 
This enables us to use any real-world photograph as our ground-truth output, and generate the corresponding input pair for training only from illumination-invariant scene properties and light simulation.
We use diverse real-world photograph collections to train our neural renderer to generate a realistic relighting result given the initial physical simulation of light.

By combining the expressiveness of PBR with the photorealism of neural rendering, our forward pipeline can render a wide variety of light sources, such as point lights, spotlights, and diffuse lights, that can be explicitly defined in the 3D space in addition to environmental illumination. 
As demonstrated in Figures~\ref{fig:teaser} and \ref{fig:qualitative_res},
our relighting pipeline can create realistic results for in-the-wild photographs for a wide variety of images under different lighting conditions.

\section{Related Work}
\label{sec:related}

Image relighting has been widely studied in the computational photography literature, offering different formulations and interactive interfaces, albeit in constrained scenarios. In this section, we provide a loose categorization of related works to provide a brief overview.

\paragraph{Supervised image relighting}

The single-image relighting literature consists of a large number of problem formulations. Due to the inherent difficulty of the problem, prior works focus on specific imagery and relighting scenarios to constrain the task.
Existing methods often make trade-offs between editability and controllability when representing the target lighting configuration. \revision{To combat the underconstrained nature of the problem, many prior works focus on the multi-view variation of the relighting problem~\cite{gardner2024sky, philip2021free, wang2023fegr, rudnev2022nerfosr}, relying on multiple captures of a given scene.} Works that tackle the single-image relighting problem constrain the domain of expected inputs, for example, focusing on portraits and objects~\cite{pandey2021total, hou2024compose, mei2023lightpainter, ren2023relightful}. The methods rely on expensive data-capture systems such as light stages.
This studio-captured data constrains their lighting representation to environmental lighting only. Similar methods have been trained using rendered data \cite{yeh2022learning, zeng2024dilightnet, jin2024gaffer, griffiths2022outcast, bharadwaj2024genlit}, but this comes at the cost of photorealism in the generated result. Attempts have been made to alleviate this data deficiency by leveraging supervision from multi-illumination~\cite{xing2024luminet, zhang2024latent, poirier2024diffusion}, or multi-view~\cite{yu2020self, philip2019multiview} data. While capturing this type of data for real images is possible, the scenes are limited to static environments in specific scenarios. 

We formulate the relighting problem as a combination of physically based rendering (PBR) and neural rendering, which allows us to train our neural renderer in a self-supervised manner where training pairs can be generated from any photograph. As a result, our method can generate realistic relighting results in a wide variety of scenes without requiring specialized dataset capture setups.

\paragraph{Inverse rendering for relighting}
A crucial step to relighting is removing the original illumination from the input scene. This step can either be performed implicitly in an end-to-end fashion \cite{bharadwaj2024genlit, zeng2024dilightnet, jin2024gaffer} or by first decomposing the image into its intrinsics in an explicit inverse rendering step \cite{xing2024luminet, choi2024scribble, pandey2021total, mei2023lightpainter, yu2020self}. 
Intrinsic decomposition and inverse rendering are long-standing problems where a variety of image formation models are adopted by different methods. 
Similar to the relighting literature, most methods focus their efforts on specific scene types such as indoor images \cite{kocsis2024iid, zeng2024rgbx, sengupta19neural, zhu2022iris, li2020inverse, zhu2022learning, luo2024diffusion}, outdoor scenes \cite{yu2021outdoor}, or portraits \cite{zhang2022phaced, wang2022face}. 
Inverse rendering alone is not enough to relight an image, as the target illumination has to be applied to the illumination-invariant representations of scene. 
Several works make use of a neural rendering step to combine intrinsic components with a lighting representation for relighting. 
The methods by~\citet{sengupta19neural} and \citet{careaga2023compositing} utilize an environment map along with estimated albedo and surface normals to re-render the input. 
While this allows them to customize environmental illumination, their use of surface normals for geometry makes it challenging to render realistic shadows. 

\paragraph{Interactive image relighting}
\citet{zeng2024rgbx} and \citet{luo2024diffusion} propose an end-to-end diffusion approach for inverse rendering and neural re-rendering, where their neural renderer takes all available intrinsics as raw input to generate the final result. 
In order to generate a novel relighting, the user can omit the shading layer and provide a natural language description of the desired lighting. 
Instead of explicit intrinsics, \citet{xing2024luminet} utilize illumination-dependent and -independent latent descriptors.
By swapping the illumination-dependent information of one image with the illumination-independent information from another scene, they demonstrate lighting transfer between two scenes.
The methods of~\citet{choi2024scribble} and~\citet{mei2023lightpainter} develop interactive approaches for indoor relighting by allowing a user to annotate the input image with scribbles to hint at the desired lighting conditions.

As is the case for \citet{choi2024scribble}, \citet{zeng2024rgbx}, and others, rendering the final image directly from intrinsic maps or latent representations makes the task of their neural renderers challenging, while also constraining their input light representation to be indirect, such as scribbles or text prompts. 
Our method, on the other hand, uses PBR to model the image formation through ray tracing. 
This makes the design of the target 3D environment explicit, enabling the user to experiment within a familiar CG environment with physically intuitive control over illumination. 
Moreover, since we model the image formation already through PBR, this simplifies the task of our neural renderer to simply bridge the gap between the PBR image and real-world appearance. 
The efficiency of our pipeline, which combines PBR and neural rendering, enables illumination editing at interactive speeds.

\section{Physically Controllable Relighting}
\label{sec:method:fwd}

\begin{figure*}
  \includegraphics[width=\linewidth]{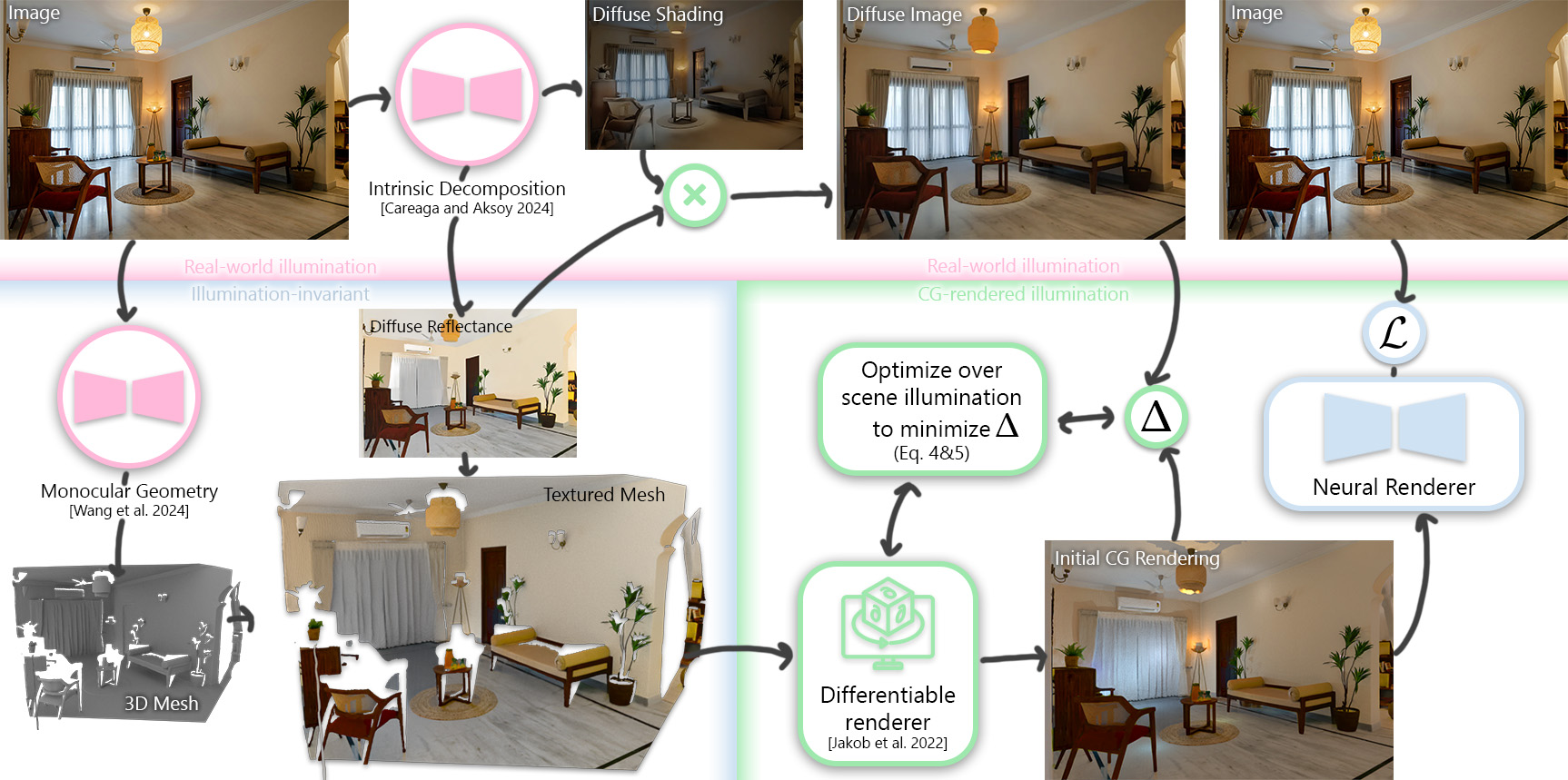}
  \caption{
    Overview of our self-supervised training process. For a given image, we first generate illumination-invariant PBR components, namely, albedo and geometry using off-the-shelf methods\cite{careaga2024colorful, wang2024moge}. We use these representations to generate a colorful 3D mesh of the scene. Using differentiable rendering implemented by Mitsuba, we recover the lighting configuration of the original scene by re-rendering it from the illumination-invariant representations. We then train a neural rendering network to recover a photorealistic rendition of the scene from the CG-rendered approximation. By removing and recovering the illumination of a single scene, we can train our neural rendering network completely with self-supervision.\\ \mbox{} \hfill \footnotesize{Image from Unsplash by Sanju Pandita.}
  }
  \label{fig:pipeline_train}
\end{figure*}
   
We introduce a single-image relighting pipeline that brings the physical control over light sources present in computer graphics pipelines to in-the-wild realistic photograph editing.
In order to allow users to control the illumination in a 3D rendering environment, our first task is to create a representation for the input photograph that can be used by physically-based rendering (PBR) engines to render the scene under customizable illumination. 
This requires an illumination-invariant 3D geometric representation of the scene.

In order to recreate in-the-wild photographs inside a rendering engine, we make use of two mid-level vision approaches that have been specifically designed for in-the-wild generalization. 
We model the scene geometry using the monocular geometry estimation method MoGe by \citet{wang2024moge}. 
MoGe estimates a geometrically accurate 3D point cloud \revision{and camera parameters}, from a single image, and has been shown to accurately represent complex scenes both indoors and outdoors. 
We estimate the diffuse reflectance for the input image using the colorful intrinsic decomposition method CID by \citet{careaga2024colorful,careaga2023intrinsic}. 
CID decomposes its input $I$ into 3 intrinsic layers using the intrinsic residual model:
\begin{equation}
    I = A \times S + R,
\label{eq:method:intrinsicmodel}
\end{equation}
where $A$ represents the diffuse reflectance or albedo, $S$ represents the diffuse RGB shading, and the residual layer $R$ includes all non-diffuse components such as specularities.
CID can generate intrinsic decompositions at high resolutions and has been shown to accurately represent the diffuse reflectance and shading colors in mixed-illumination environments both indoors and outdoors.

We generate a mesh using a 3D point cloud from MoGe by triangulating the positions of neighboring pixels. 
We combine our mesh with the diffuse reflectance colors from CID to create a textured mesh that can readily be used by rendering engines. 
Note that both the geometry and the reflectance are illumination-invariant properties. 
We load this illumination-invariant 3D scene representation into a rendering engine in which the user can define new light sources in the scene.

Although this combination of MoGe and CID allows us to generate 3D scene representation for in-the-wild photographs, it is still an incomplete representation lacking some characteristics that a full CG pipeline requires for photorealistic rendering.
One missing component is a result of the monocular view of the scene where only the surfaces visible from the camera can be reconstructed. 
This results in an incomplete mesh of the scene as shown in Figure~\ref{fig:pipeline_fwd}, where occluded regions are omitted in the 3D point cloud estimated by MoGe. 
Due to the lack of neighboring pixels at depth discontinuities, the generated mesh also fails to represent surfaces close to occlusion boundaries. 
Similarly, CID provides the illumination-invariant diffuse reflectance but combines all non-diffuse effects in the residual layer which cannot be used to assign non-diffuse surface properties. 
While diffuse reflectance can still successfully simulate secondary illumination coming from colored surfaces, it fails to represent specular surfaces or refractive objects. 
Due to a lack of reliable BRDF estimation methods in the literature for in-the-wild photographs, our 3D representation only allows us to render a diffuse image.

In our relighting pipeline, we task our neural renderer to compensate for the shortcomings of our 3D textured mesh coming from monocular views to generate photorealistic final results. 
As outlined in Figure~\ref{fig:pipeline_fwd}, we render an initial image via PBR using tools such as Blender or Mitsuba that allow the user to customize the illumination environment. 
This initial rendering serves as the input to our neural renderer which generates the final realistic image. 

Our neural renderer needs to be trained on real-world ground-truth data in order to render a realistic result complete with non-diffuse effects. 
In the next section, we outline the self-supervised training pipeline that we formulate using PBR and differentiable rendering.

\section{Self-supervised Neural Rendering}
\label{sec:method:train}

The main task of our neural renderer is to generate a realistic image that reflects the appearance of real-world photographs given an initial PBR render. 
To get our neural renderer to successfully model real-world appearance, we need to train it with real-world photographs as ground-truth. 
In this section, we detail our method of generating input PBR rendering and output real-world appearance pairs for training using only a set of photographs.

In order to use a photograph $I$ as the ground-truth appearance, we need to generate a PBR version of the scene under the existing illumination. 
Similar to our forward pipeline outlined in Section~\ref{sec:method:fwd}, we use MoGe \cite{wang2024moge} and CID \cite{careaga2024colorful} to create a textured mesh $M$ for $I$. 
Given the geometry and diffuse reflectance, we are only missing the light sources to create our initial rendering using ray-tracing. 
We now detail our optimization over the 3D illuminating environment that will act as a stand-in for user-defined lighting during training.

\subsection{Optimization via Differentiable Rendering}
\label{sec:method:optimize}

Given an input image $I$ and the illumination-invariant textured mesh $M$, our goal is to use a ray tracer $\mathsf{pbr}$ to render the best approximation of $I$ that our incomplete monocular estimation of $M$ allows. 
For this purpose, we formulate the 3D lighting environment estimation as an optimization problem. 

\subsubsection{Target variable} 
As noted in Section~\ref{sec:method:fwd}, our 3D representation $M$ only models diffuse reflectance. 
This makes it challenging to use the original image $I$ as the target variable of our optimization, as $I$ includes non-diffuse effects. 
This is because non-diffuse effects such as specularities typically result in very bright regions in the image, skewing our optimization. 
Instead, we use the diffuse image as our target variable during optimization. 
CID allows us to obtain the diffuse image $D$ using the diffuse reflectance $A$ and diffuse shading $S$ as defined in Equation~\ref{eq:method:intrinsicmodel}:
\begin{equation}
    D = A \times S.
\end{equation}

\subsubsection{Unknown variables}
Without loss of generality, we represent the unknown 3D lighting environment ${\Psi}$ as a combination of an environmental illumination $E$ and a set of point light sources $\mathcal{P} = \{ \vec{p}_i | i \in\{1,2,...,K\}\}$:
\begin{equation}
    \Psi= \{ E, \mathcal{P} \}, 
\end{equation}
where $E>0$ is a high dynamic range image (HDRI) map. 
The point light sources $\vec{p}_i \in \mathbb{R}^6$ are defined as the concatenation of non-negative RGB intensities and unconstrained 3D locations.

\subsubsection{Objective function}
Our objective is to reconstruct the diffuse image $D$ as closely as possible using the input textured mesh $M$ in a PBR environment by optimizing over the 3D lighting environment ${\Psi}$.
Formally, our objective function can be defined as:
\begin{equation}
    e(D,M,\Psi) = \sum_{\forall i \in \mathcal{V}} \sum_{\{r,g,b\}} \left(  D_i - \mathsf{pbr}(M,\Psi)_i \right)^2,
\label{eq:method:objective}
\end{equation}
where $\mathsf{pbr}(M,\Psi)$ is the physically-based rendering operation and $\mathcal{V}$ is the set of valid pixels in our rendering, excluding holes coming from the monocular geometry estimation.

\subsubsection{Optimization}
We fit the 3D lighting environment $\Psi$ to best match the original diffuse image through nonlinear optimization:
\begin{equation}
    \Psi^* = \argmin_\Psi e(D,M,\Psi)
\label{eq:method:argmin}
\end{equation}
and use the optimized lighting $\Psi^*$ to create our initial diffuse rendering $\tilde{D} = \mathsf{pbr}(M,\Psi^*)$ that will serve as the input to our neural renderer during training.
We solve this nonlinear minimization problem using the differentiable renderer Mitsuba 3 \cite{jakob2022mitsuba3} and the gradient-based Adam optimizer \cite{adam}. 

\subsubsection{Parameters}
We define our environmental illumination $E$ as a $128\times 256$ HDRI RGB map and define $K=16$ point light sources. 
This results in a 98400-dimensional optimization over $\Psi$, $16\times 6 = 98$ variables representing the 16 point light sources and the rest coming from $E$.
We determine the resolution for $\mathsf{pbr}(M,\Psi)$ by setting the longer dimension of the image to $512$ pixels while maintaining the original aspect ratio. 
We resize $D$ to compute $e$ at this resolution. 
We initialize $E$ as a constant gray image at $0.5$, set the RGB intensities of $\vec{p}_i$ at the same value, and position the 16 point lights in a $4 \times 4$ regular grid at the center of the scene. 
We generate $\mathsf{pbr}(M,\Psi)$ with ray tracing in Mitsuba using a bounce depth of 3 representing up to the first-bounce indirect illumination with 16 samples per pixel for Monte-Carlo rendering.
We use the initial learning rate of $0.01$ for Adam.
It takes on average 20 seconds per image to optimize for $\Psi$ on an NVIDIA A40 GPU.

\subsubsection{Discussion}
\label{sec:method:optimize:discussion}
Although we are optimizing over 3D lighting parameters, it should be noted that our main goal with this optimization is to reconstruct $D$ as faithfully as possible in our simplified PBR environment. 
In order to transfer user-defined illumination conditions into a photorealistic image, we expect the neural renderer to remain faithful to its input rendering in terms of diffuse lighting. 
This is to simplify the task of our neural renderer into only modeling the domain gap between physically-based diffuse rendering and real-world appearance.

Our choice of a low-resolution target HDRI and modeling non-environmental lighting only as a small set of point lights are motivated by the limitations of our 3D representation of the scene and the nonlinearity of our optimization problem. 
We generate our textured mesh $M$ used as the implicit variable in our optimization using computer vision methods MoGe \cite{wang2024moge} and CID~\cite{careaga2024colorful}.
While MoGe can accurately model complex geometries in the wild, it falls short in creating a high-precision geometry at high resolutions. 
Combined with the inherent incompleteness of monocular geometry in occluded areas and occlusion boundaries, our mesh $M$ only roughly approximates the full 3D geometry.
CID can accurately generate high-resolution reflectance in the wild, but it can only model the diffuse reflectance as a variable that can be used for PBR. 
Diffuse-only rendering can still model the colorful illumination in the scene including multi-bounce indirect light, but makes it challenging to infer high-resolution environmental lighting due to the lack of direct reflections and specularities.

The computation of our objective function involves rendering the scene via PBR. 
As a result, the optimization requires a differentiable renderer. 
Mitsuba 3 \cite{jakob2022mitsuba3} is a differentiable rendering system that makes it possible to compute the partial derivatives for many independent variables. 
Although Mitsuba supports various types of illumination sources, we find that the combination of an environment and point light sources allows our optimized lighting to approximate many different illumination conditions, without creating an unstable optimization process.

\begin{figure}
  \includegraphics[width=0.9\linewidth]{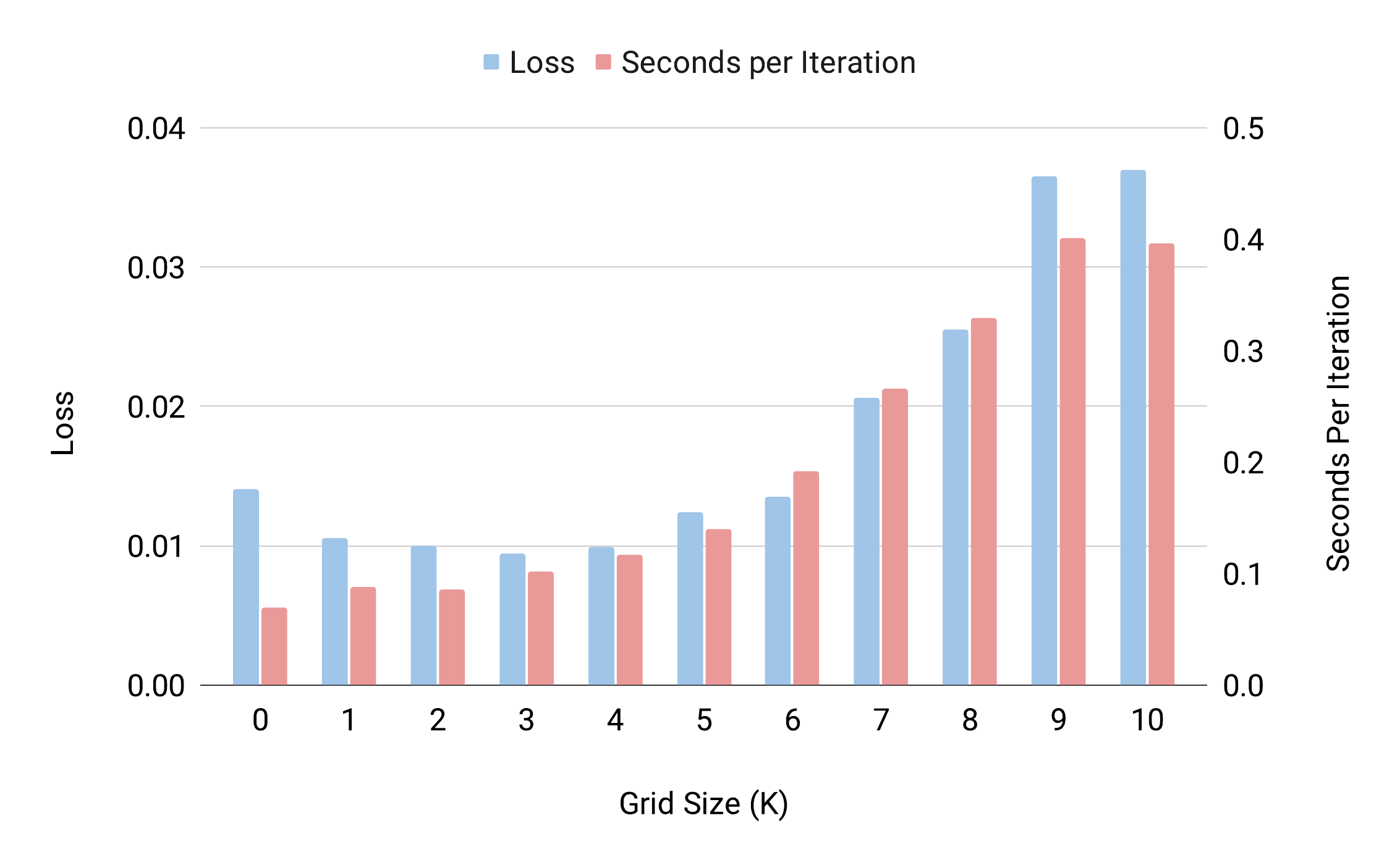}\vspace{-0.4cm}
  \caption{Optimization runtime and performance as a function of the number of point lights. We find that using 9-16 point lights creates a stable optimization problem, improves over only environment lighting, and is efficient.}
  \label{fig:opt_analysis}
\end{figure}
\begin{figure*}
  \includegraphics[width=\linewidth]{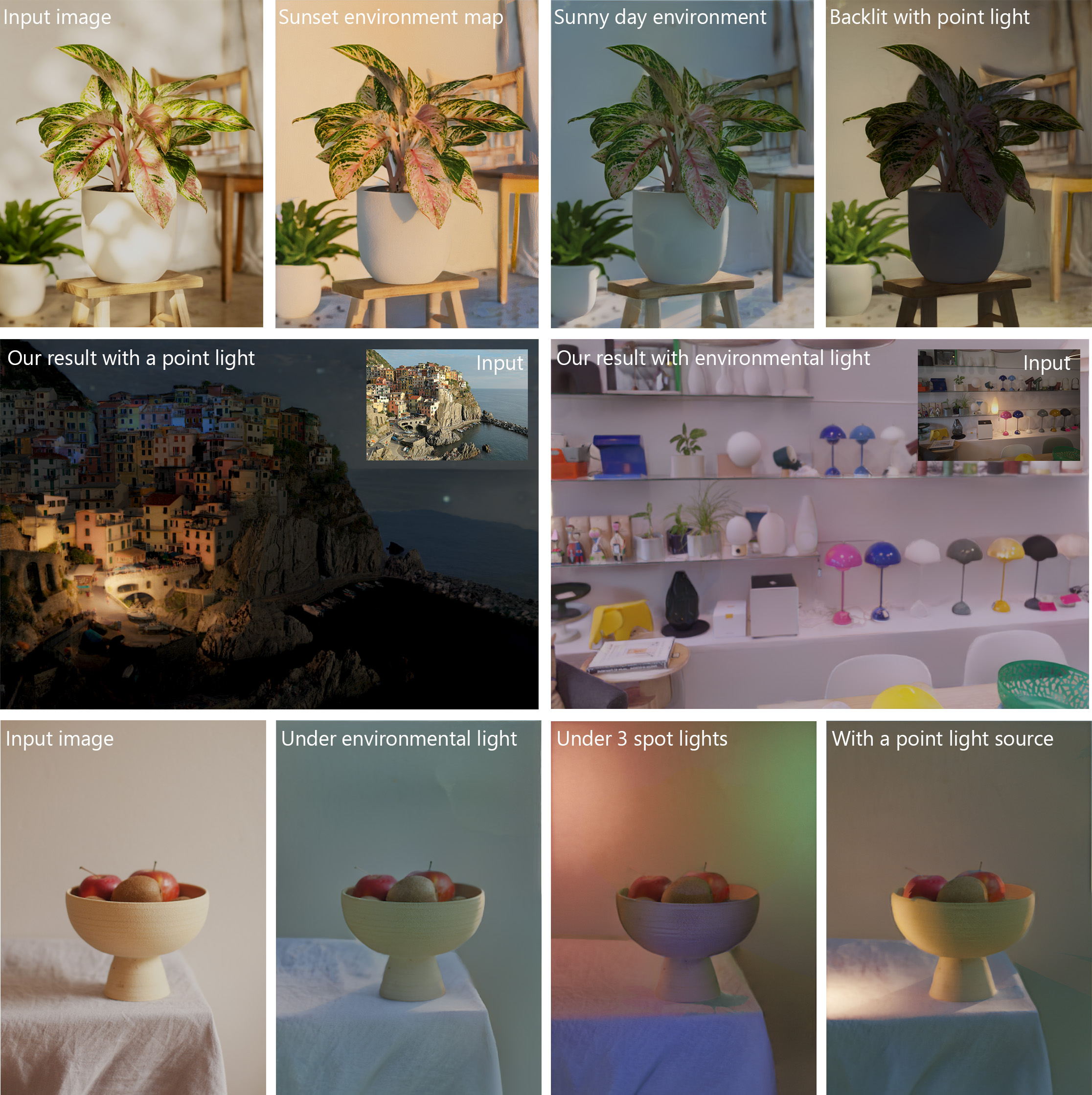}
  \caption{Our method can generate unconstrained, full-controllable relit images for a wide variety of scenes. By defining the lighting configuration in 3D, users can easily add both global and local lighting effects. Our neural renderer then converts CG-rendered approximations into realistic imagery. \\
  \mbox{} \hfill \footnotesize{Images from Unsplash by Feey (plant), Maksim Shutov (Cinque Terre), and Suzanne Boureau (fruit)}}
  \vspace{2cm}
  \label{fig:qualitative_res}
\end{figure*}

We define our optimization over the HDRI environment intensities, point light source intensities, and point light source locations. 
With its $128\times 256$ resolution, the HDRI map constitutes a majority of the optimized variables. 
However, thanks to the linearity and the superposition properties of illumination, both HDRI and point light intensities create little nonlinearity in the PBR-based gradients, making the high-dimensionality of the optimization process more manageable.
The 3D locations of the point light sources, however, make our objective function $e$ highly nonlinear. 
This nonlinearity comes from the complex interactions between light sources and the geometry, such as cast shadows.
As a result, our optimization gets prone to local minima as we increase $K$ as Figure~\ref{fig:opt_analysis} shows.
Our choice of using a $4\times 4$ grid of $16$ point lights reflects the best reconstruction performance, balancing stability and expressivity.

Despite these limitations, as we show in Figure~\ref{fig:pipeline_train} and in the supplementary, our 3D lighting representation ${\Psi}$ is capable of simulating close approximations to the diffuse image $D$ in a variety of scenes and real-world illumination conditions. 
Although ${\Psi}$ allows us to render input images for the training of our neural renderer, it is not intended to be an accurate modeling of real-world light sources.

\subsection{Neural renderer}

The final step in our pipeline is the feed-forward neural renderer (NR) that models the gap between the initial rendering $\tilde{D}$ and real-world appearance. 
Our lighting optimization in Section~\ref{sec:method:optimize} allows us to use any real-world photograph for training. 
Given the image $I$, we first generate an illumination-invariant 3D representation of the scene, which is then rendered under CG lighting. 
We use the diffuse image $D$ as the target variable to generate the closest PBR approximation to the original illumination in the environment. 
$I$ itself serves as the ground-truth for real-world appearance.
We use physical modeling to generate the illumination-dependent inputs to NR, effectively isolating it from existing illumination in $I$. 
Illumination-variant variables $I$ and $D$ are used only as target variables for either NR or PBR.
Our self-supervised training pipeline is outlined in Figure~\ref{fig:pipeline_train}.

\subsubsection{Inputs and losses}
Our initial rendering $\tilde{D}$ reflects the target illumination conditions, but lacks details due to its incomplete and smooth geometry. 
In order to allow our network to remain faithful to the original scene content, we also provide the full-resolution diffuse reflectance $A$ as input. 
As noted in Section~\ref{sec:method:optimize:discussion}, the incomplete geometry results in missing pixels in the PBR result. 
We apply a low-level hole filling to clean $\tilde{D}$ of high-frequency artifacts and provide a binary mask of invalid pixels, $\mathcal{V}^c$, also as input. 
We concatenate $\tilde{D}$, $A$, and $\mathcal{V}^c$ into a 7-channel map to serve as the input to NR.

The output of NR, $\tilde{I}$, is defined as the realistic relighting result in linear RGB.
Using the original image $I$ as ground-truth, we define our loss as a combination of the mean-squared error and the commonly used multi-scale gradient loss \cite{li2018mega}:
\begin{equation}
    \mathcal{L} = MSE(I, \tilde{I}) + \sum_m MSE(\nabla {I}^{m}, \nabla \tilde I^{m}),
\label{eq:method:networkloss}
\end{equation}
where $\nabla I^{m}$ represents the gradient of $I$ at scale $m$.

\subsubsection{Training dataset}
We derive both the implicit variable $M$ and the target variable $D$ in our optimization in Equation~\ref{eq:method:argmin} directly from the input image $I$, which also serves as the ground-truth real-world appearance for our neural renderer. 
This allows us to use any set of real-world photographs to generate training data.
We use the publicly available raw photograph collections RAISE \cite{raise}, MIT 5k \cite{fivek}, PPR 10k \cite{jie2021PPR10K}, and LSMI \cite{kim2021lsmi}.
This large collection of in-the-wild photographs offers a wide variety of training images that are complex in terms of scene geometry, indoor and outdoor environments, direct and environmental real-world illumination conditions, and objects and materials.
In order for our neural renderer to model the appearance in a higher dynamic range, we choose to utilize raw photographs for training instead of \texttt{jpg} collections. 
Note that since the higher dynamic range comes from PBR in our forward pipeline, this choice of training on raw images does not result in a raw input requirement in our forward pipeline.

We pre-process the entire dataset to generate $(\tilde{D},I)$ pairs prior to training.
Although our optimization can successfully approximate diverse real-world lighting conditions, it may fail in some scenes with visible shadows in the image that are cast from out-of-view objects. 
We filter out such images that our optimization can not reliably reconstruct using a $2.5D$ monocular geometry using the minimized objective function $e(D,M,\phi^*)$, which directly reflect the reconstruction accuracy. 
We eliminate $15\%$ of the training pairs with the highest $e(D,M,\phi^*)$ to ensure consistency between all $\tilde{D} - I$ pairs used during training.

\subsubsection{Architecture and Training}
For our neural rendering network, we employ a commonly used encoder-decoder architecture from Midas \cite{ranftl2020towards}. We replace the encoder with the anti-aliased version by \citet{zhang2019shiftinvar}. The network outputs the linear relit image directly, using a ReLU activation to clip negative values. We train the network using the Adam optimizer with a learning rate of $10^{-5}$. Each batch has 8 images at 384x384 resolution, and the network is trained for approximately 2 million iterations, which takes 10 days on a single NVIDIA A40 GPU. We perform random cropping and flipping as augmentation.

\section{Experiments}
\label{sec:experiments}

\begin{figure*}
  \includegraphics[width=\linewidth]{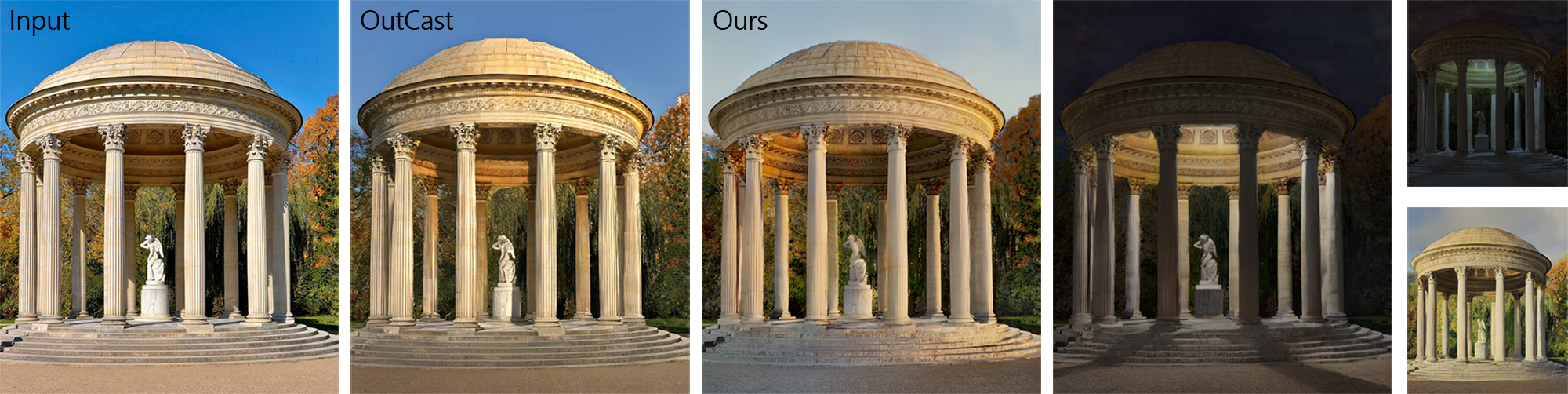}
  \caption{The method of~\cite{griffiths2022outcast} performs state-of-the-art outdoor scene relighting. Their method can only represent very specific lighting configurations as it models the position of the sun in the sky. Our method is able to perform this same task while also being unconstrained, allowing for precisely controlled relighting of the scene with local lighting effects. \hfill \footnotesize{Image provided by \citet{griffiths2022outcast}}}
  \label{fig:outcast_comp}
\end{figure*}
\begin{figure*}
  \includegraphics[width=\linewidth]{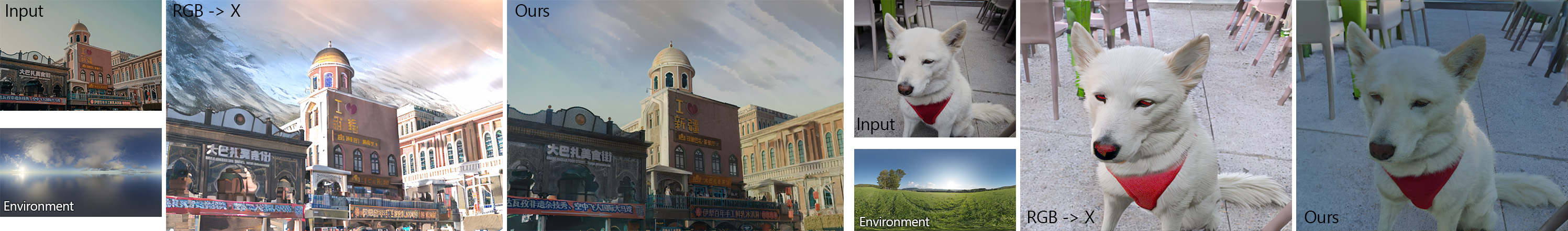}
  \caption{We compare our results to the X$\Rightarrow$ RGB portion of the pipeline from \cite{zeng2024rgbx}. Since the method relies on an input shading map to synthesize the image, we feed it our rendered shading after inpainting the missing regions. The model is able to utilize this shading but does not generate results consistent with the input shading. Additionally, much of the image content is altered in the resulting relit image. \\ \mbox{} \hfill \footnotesize{Images from Unsplash by Lywin (bazaar) and the HDR+ Dataset~\cite{hasinoff2016burst} (dog) .} }
  \label{fig:rgbx_comp}
\end{figure*}

In this section, we provide qualitative and runtime comparisons to representative state-of-the-art scene relighting methods and discuss the shortcomings of each relative to our method. We also provide quantitative comparisons in the form of a user study and a numerical experiment on the BigTime dataset in the supplementary material.

\subsection{Qualitative Results}

We show a wide variety of in-the-wild relighting examples in Figures~\ref{fig:teaser}, \ref{fig:pipeline_fwd}, and \ref{fig:qualitative_res}. 
Our PBR-based simulation of the illumination allows us to insert physically realistic lights such as the headlights on the car in Figure~\ref{fig:teaser}, spotlights in Figure~\ref{fig:pipeline_fwd} and in the fruit bowl example in Figure~\ref{fig:qualitative_res} as well as point light sources as shown in  Figures~\ref{fig:pipeline_fwd} and \ref{fig:qualitative_res}.
Our physical reconstruction of the scene enables extreme changes such as day-to-night or night-to-day conversion, even under existing complex night-time lighting (Figure~\ref{fig:teaser}) or strong sunlight (Figure~\ref{fig:outcast_comp}).
We further provide video examples with moving light sources in the supplementary video. No prior method is capable of performing a fully controllable relighting with light sources defined in a 3D environment. Therefore, we show some conceptual comparisons to existing state-of-the-art methods.

\paragraph{OutCast}
Figure~\ref{fig:outcast_comp} shows a conceptual comparison to a result from the method of~\citet{griffiths2022outcast}. The method is conditioned by a sun position, meaning there is a specific set of possible lighting configurations that they are able to model. We can accurately generate relit images under this setting with an environment map, but we can also generate any other configuration of environmental and local lights, as shown in the additional examples.

\begin{figure*}
  \includegraphics[width=\linewidth]{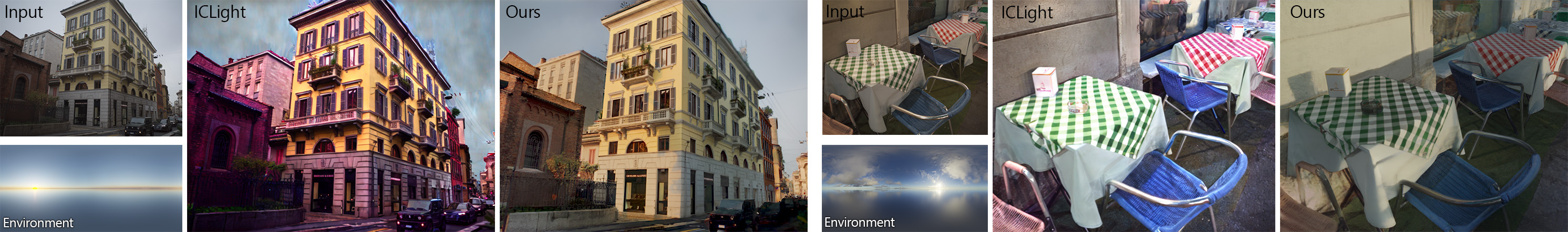}
  \caption{\revision{Comparison to ICLight~\cite{zhang2025scaling}. We condition their relighting diffusion model using the same initial CG render that we provide to our model. Their method can generate a reasonably realistic result, but fails to remove the illumination from the input image. Our method properly uses the input signal to achieve the desired environmental lighting provided.}}
  \label{fig:iclight_comp}
\end{figure*}

\paragraph{RGB$\Leftrightarrow$X} In Figure~\ref{fig:rgbx_comp} we show comparisons to relighting results of RGB$\Leftrightarrow$X~\cite{zeng2024rgbx}. Their re-rendering network requires estimated shading for the target re-rendering, or a natural text description in its place. To compare one-to-one with our approach, we feed their diffusion model our rendered shading layer as the \emph{irradiance} channel. There model is able to utilize this guidance but is unable to model the realism of natural photographs due to the lack of real-world data in their training distribution. Additionally, the model expects a clipped shading layer in the $[0, 1]$ range, therefore, it can not represent the full dynamic range of the target relighting. Our method, on the other hand, is able to generate physically-accurate lighting effects using the rendered starting point. In the right example, we can see that our model is able to relight the dog with accurate green illumination from the grassy environment, and a bright spot from the sun behind the scene.
\revision{
\paragraph{ICLight}
In Figure~\ref{fig:iclight_comp} we compare to ICLight \cite{zhang2025scaling}, which is a diffusion model meant to harmonize a foreground object or portrait with a given background scene. The model can be used as a general relighting method by feeding the input image as the foreground, and a lighting condition image as the background. We use our CG render as the lighting condition. We find that the method is oftentimes not able to remove the original lighting of the input scene since their pipeline does not explicitly take advantage of illumination-invariant scene representations such as albedo. Our method, on the other hand, can apply completely novel illuminating environments without residual shading effects from the input image.}

\begin{figure*}
  \includegraphics[width=\linewidth]{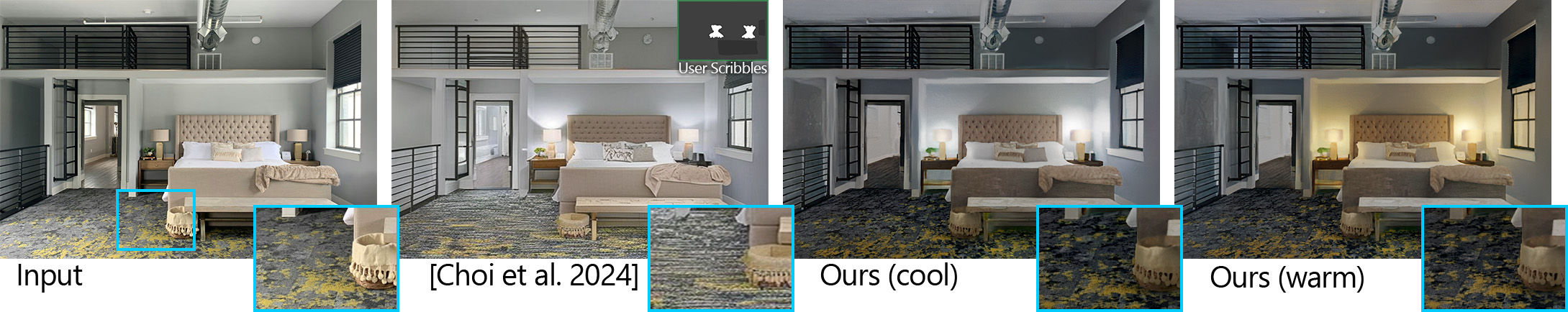}
  \caption{
    The method of \citet{choi2024scribble} uses user-scribbles to condition indoor relighting. Rather than rely on indirect conditioning in the image space, we allow the user to explicitly place light sources in the 3D scene. This means we can accurately model resulting lighting effects, such as the light bouncing off the wall. ScribbleLight requires the user to explicitly annotate these effects. Additionally, the result from ScribbleLight contains altered content due to the generative modeling of the problem (inset), whereas our result maintains the identity of the original scene. \hfill \footnotesize{Image from Unsplash by Adam Winger}}
  \label{fig:scribble_comp}
\end{figure*}

\paragraph{ScribbleLight}
In Figure~\ref{fig:scribble_comp}, we show a comparison with ScribbleLight~\cite{choi2024scribble}, which makes use of estimated intrinsic components and user-drawn scribbles to condition a relighting diffusion model. Although their model can easily be conditioned, the 2D nature of the scribbles limits the expressiveness of the control. Users have to manually annotate that they want certain areas darkened or lightened. Our method, on the other hand, just requires the user to place a 3D light source where they want it in the scene. We can then infer the side effects of the light, like saturated regions on the wall, and darkening at the foot of the bed. Additionally, the generative nature of the ScribbleNet pipeline results in altered scene content as shown in the inset. Our method remains faithful to the original image, which is crucial for photo relighting.

\subsection{Runtime Comparison}
One of our framework's main strengths when compared to existing approaches is efficiency. While we do require estimated PBR quantities for a given scene, we use feed-forward methods to generate these components, taking only 2 seconds for a 512x512 image. After this initial preprocessing, the image can be relit multiple times. For each relighting, we render the generated mesh and send it through our neural renderer. With 512 spp rendering, this whole process can be carried out in about 0.7 seconds. Since recent relighting works build off of Stable Diffusion, their pipelines require an expensive sampling step. For a 512x512 image, the re-rendering portion of RGB$\Leftrightarrow$X takes about 6 seconds to generate the output. The authors of LumiNet~\cite{xing2024luminet} cite similar timings for their method. Our efficiency allows us to perform interactive editing with responsive user feedback and can readily be integrated into PBR frameworks such as Blender.
We demonstrate our pipeline in action in a simple interactive web-based application in the supplementary video.


\section{Limitations}

As we start from a single image, we can only reliably work with a $2.5D$ monocular geometry, which can not represent occluded areas and out-of-view geometry. 
For many common light configurations, any artifacts coming from this representation are fixed by our neural renderer. 
However, placing the light sources in extreme places, such as behind the scene,e may result in unrealistic cast shadows as the light can leak into the scene from the missing geometry. 

We do PBR using the diffuse reflectance and rely on our neural renderer to create non-diffuse lighting effects in the scene. 
While this works for a variety of materials and scenes, one challenging case is portrait images.
Human skin and hair require highly complex models to allow a realistic simulation of light, and 3D reconstruction of faces and hair also requires very high precision. 
With the estimated diffuse reflectance and an in-the-wild geometry estimator, most of the intricacies of the portrait are lost to our light simulation. 
With a dedicated material model and specialized geometry estimators, however, it is possible to utilize our self-supervised system design for portrait images in the future.

\section{Conclusion}

Our pipeline brings the physical control over the light sources, that has only been available in computer graphics pipelines, to image relighting. 
To achieve this, we make use of state-of-the-art mid-level computer vision methods, differentiable physically based rendering, and neural rendering that we train with a self-supervised strategy.
This allows us to combine the in-the-wild generalization of modern computer vision methods, the expressivity and customizability of computer graphics pipelines, and the real-world appearance modeling power of neural networks in a single computational photography pipeline.
With the recent speed at which in-the-wild high-resolution computer vision methods have been progressing and the ongoing development of advanced PBR tools like Mitsuba 3, we expect to see more opportunities to integrate computer graphics and computer vision methodologies for computational photography applications.

\begin{acks}
We would like to thank Mahdi Miangoleh, Sebastian Dille, and Samuel Atunes Miranda for their help with Blender, the supplementary video, and the interactive interface. We would also like to thank Jun Myeong Choi for promptly providing results for their method. We acknowledge the support of the Natural Sciences and Engineering Research Council of Canada (NSERC), [RGPIN-2020-05375].
\end{acks}

\bibliographystyle{ACM-Reference-Format}
\bibliography{references}

\end{document}